\documentclass[pdflatex]{pasj02}
\usepackage[switch,mathlines]{lineno}
\usepackage{url}
\usepackage{mathcomp}
\usepackage{tabularx}
\usepackage{makecell}
\usepackage{adjustbox}

\Received{$\langle$reception date$\rangle$}
\Accepted{$\langle$acception date$\rangle$}
\Published{$\langle$publication date$\rangle$}

\newcommand{\Kc}{K25}
\newcommand{\Kcs}{K25 }

\begin{document}

\title{Orbital inclination estimates for overcontact binaries using the derivatives of light curves}

\author{Shinjirou Kouzuma}

\altaffiltext{}{Faculty of Liberal Arts and Sciences, Chukyo University, 101-2 Yagoto-honmachi, Showa-ku, Nagoya, Aichi 466-8666, Japan}
\email{skouzuma@lets.chukyo-u.ac.jp}

\KeyWords{binaries: close --- binaries: eclipsing --- methods: data analysis}

\maketitle

\begin{abstract}
The orbital inclination of an eclipsing binary is generally determined through light curve analysis. 
Binary parameters in the light curve analysis are typically constrained through the use of optimization and sampling techniques. 
We propose a new simple method, based on the derivatives of light curves, for estimating the orbital inclinations of overcontact systems. 
Our sample consists of 89,670 synthetic light curves for overcontact binaries, covering a parameter space typical of overcontact systems. 
We classified the sample light curves on the basis of a recently proposed classification scheme: DP, SPp, SPb, SPf, and SPs types. 
For each type, we found that the orbital inclination is closely associated either with the time interval between local extrema in the derivatives of light curves or with the depth of the local minimum at phase 0.5 in the second derivative. 
Using regression analysis of the identified associations, we developed empirical formulae to estimate the orbital inclinations for each type of light curve. 
We also provide the associated uncertainties for the estimated inclinations. 
Application of the proposed method to real overcontact binary data demonstrated that our method can reasonably estimate both the inclinations and their uncertainties. 
\end{abstract}

\section{Introduction}
The orbital inclination is defined as the angle between the plane of the sky and the orbital plane of a binary system, conventionally taken to be between $0\tcdegree$ and $90\tcdegree$. 
In eclipsing binaries, the inclination strongly influences the shape of a light curve (LC). 
When the inclination is close to $90\tcdegree$, the binary exhibits total-annular eclipses; 
as long as the eclipses remain total-annular, the eclipse depth is almost independent of the inclination angle. 
If the eclipses exhibit a flat-bottomed shape, the width of the flat bottom should gradually decrease with decreasing inclination. 
As the inclination decreases and thus the eclipse shifts away from the center, the eclipse becomes partial. 
In such situations, both the depth and width of the eclipses in the LC tend to decrease further as the inclination becomes lower. 
If the inclination is sufficiently low, eclipses disappear from the LC, and the system will no longer be observed as an eclipsing binary. 

The inclination is essential for determining the absolute parameters of a binary system. 
The semi-major axis and component stars' masses derived from radial velocity curves are expressed as functions of the inclination (e.g., \cite{Hilditch2001-icbs}). 
This relationship indicates that determining the inclination is necessary to obtain such key parameters of the binary system. 

Determining the inclination is generally based on LC analysis, in which theoretical LCs are synthesized using codes such as Wilson-Devinney \citep{Wilson1971-ApJ}. 
In this analysis, the binary parameters are determined by modeling the LC through an iterative process. 
Optimization techniques, including the least-squares method, are often employed to determine binary parameters. 
Alternatively, sampling methods, such as Markov Chain Monte Carlo (MCMC; see e.g., \cite{Hogg2018-ApJS}), provide robust posteriors and uncertainties of parameters \citep{Conroy2020-ApJS}. 
Providing an appropriate initial parameter space is crucial for obtaining reliable results using these approaches. 

Overcontact binaries are notable systems in which both component stars overfill their Roche lobes. 
This configuration allows us to estimate the photometric mass ratio directly from the LC, even without radial velocity curves \citep{Wilson1994-PASP}. 
Whereas the photometric mass ratio can be accurately determined for overcontact systems with total-annular eclipses, its reliability is likely to significantly decrease for systems with partial eclipses \citep{Terrell2005-ApSS, Terrell2022-Galax}. 
Many studies have estimated the photometric mass ratios of overcontact systems by modeling their LCs with an iterative process (e.g., see the catalog by \cite{Latkovic2021-ApJS}). 

\citet{Rucinski1993-PASP} proposed a method for estimating essential parameters of overcontact systems based on Fourier coefficients of their LCs. 
This method does not require an iterative process and is applicable to a large number of systems \citep{Selam2004-AA}. 
However, it is sensitive to the presence of third light, which reduces the amplitudes of the variations in the LC. 
\citet{Hambalek2013-CoSka} showed that the degeneracy of a solution derived solely from a LC is inversely proportional to the amplitude of the LC. 
Therefore, in systems with unknown or poorly constrained third light, solutions based solely on LCs can become highly degenerate and may lead to biased or erroneous parameter estimates. 

Recent studies proposed a new method to estimate mass ratios using the derivatives of LCs \citep{Kouzuma2023-ApJ, Kouzuma2025-PASJ-1323}. 
This method mainly requires the derivatives of LCs and the measurement of the time interval between two extrema, without requiring an iterative process. 
This method also provides reasonable uncertainties for the estimated mass ratios. 
A straightforward method to estimate binary parameters, including the mass ratio, is expected to be helpful for a general understanding of individual systems; it leads to a groundwork for more detailed analyses, such as the initial estimation of parameter space for the LC analysis. 
Furthermore, obtaining binary parameter estimates with reliable uncertainties allows statistical analyses to be conducted with reasonable reliability, especially when applied to large datasets. 

This paper presents a simple method for estimating the orbital inclinations of overcontact binaries using the derivatives of LCs. 
Section \ref{sec:data} introduces the synthetic LCs of overcontact binaries for identifying key values that are closely associated with the inclination. 
Section \ref{sec:method} describes the empirical formulae to estimate inclinations from the identified key values, along with the estimation of their uncertainties. 
In section \ref{sec:application}, we introduce real binary data to examine the effectiveness of the proposed methods. 
Section \ref{sec:result} presents its results and discusses their effectiveness. 
Finally, section \ref{sec:summary} provides a summary.

\section{Data}\label{sec:data}
In this study, we employed the synthetic LC dataset for overcontact eclipsing binaries from \citet{Kouzuma2025-PASJ} (hereafter \Kc), which was generated using the PHOEBE 2.4 code \citep{Prsa2016-ApJS, Conroy2020-ApJS}. 
This dataset includes a total of 89670 LCs synthesized across a broad parameter space that covers typical ranges of overcontact binary parameters. 
The model parameters cover mass ratios ($q=M_\mathrm{s}/M_\mathrm{p}$) ranging from $0.05$ to $0.95$ (in steps of 0.1), orbital inclinations ($i$) from $30\tcdegree$ to $90\tcdegree$ ($1\tcdegree$), fill-out factors ($f$) from $0.2$ to $0.8$ ($0.3$), and stellar temperatures ($T_\mathrm{p}$ and $T_\mathrm{s}$) from $4000$ to $10000$ K ($1000$ K). 
Here, `p' and `s' denote the primary and secondary, respectively; the primary star is defined as the more massive component of the binary. 
The gravity-darkening coefficient was set either to 0.32 or 1, depending on whether the star's temperature was lower or higher than 6600 K, respectively. 
For limb-darkening, we adopted coefficients computed by PHOEBE on the basis of PHOENIX stellar atmosphere models. 

These synthetic LCs were binned into 100 phase intervals, and their numerical derivatives with respect to time were computed up to the fourth order. 
They were also classified into five types on the basis of \Kcs (see also \cite{Kouzuma2023-ApJ} for DP): DP, SPp, SPb, SPf, and SPs types. 
We attempted to explore a simple method for estimating the orbital inclination for each of the five types.

\begin{table*}
  \tbl{Summary of the key values ($W_i$), empirical formulae ($i_\text{est}$), standard deviations ($\sigma_{W_i}$), uncertainties of $W_i$ ($\delta W_i$), and uncertainties of the estimated inclinations ($\delta i_\text{est}$) for estimating the orbital inclinations of DP, SPp, SPb, SPf, and SPs systems. }{%
	\begin{adjustbox}{width=\textwidth}
  \begin{tabular}{l>{$}l<{$}>{$}l<{$}>{$}l<{$}>{$}l<{$}>{$}l<{$}>{$}l<{$}}
      \hline
      LC type & W_i & i_\text{est} & \sigma_{W_i}  & \delta W_i & \delta i_\text{est}  \\ 
      \hline
	  DP & \displaystyle \frac{1}{2} \left(\frac{P}{t'_{41}-t'_{32}}+\frac{P}{t_{32}-t_{41}} \right) & 0.683 W_{i, \text{DP}} + 40.42 & 6.526  & \displaystyle \frac{P}{2\sqrt{2}} \left(\frac{1}{d_\text{DP}'^4}+\frac{1}{d_\text{DP}^4}\right)^\frac{1}{2} |d_\text{DP}-d'_\text{DP}| & \displaystyle 0.683 \sqrt{\sigma_{W_i, \text{DP}}^2 + \left(\delta W_{i, \text{DP}} \right)^2} \\ \\
	  SPp & \displaystyle \frac{1}{2} \left(\frac{P}{t_{32}-t'_{11}}+\frac{P}{t_{11}-t'_{32}} \right) & 16.17 \ln (W_{i, \text{SPp}}-5.600) + 54.40 & -6.05\cdot 10^{-3} i_{\text{est}}+0.718  & \displaystyle \frac{P}{2\sqrt{2}} \left(\frac{1}{d_\text{SPp}'^4}+\frac{1}{d_\text{SPp}^4}\right)^\frac{1}{2} |d_\text{SPp}-d'_\text{SPp}| & \displaystyle 16.17 \sqrt{\sigma_{W_i, \text{SPp}}^2 + \left(\frac{\delta W_{i, \text{SPp}}}{W_{i, \text{SPp}}-5.600}\right)^2} \\ \\
      SPb & \displaystyle f_{\text{n}, 0.5} & 78.87 W_{i, \text{SPb}} + 91.20 & -1.91\cdot 10^{-3} i_{\text{est}}+0.195  & \delta f_{\text{n}, 0.5} & \displaystyle 78.87 \sqrt{\sigma_{W_i, \text{SPb}}^2 + \left( \delta W_{i, \text{SPb}} \right)^2} \\ \\
	  SPf & \displaystyle \frac{P}{t_{11}-t'_{11}} & 7.384 \ln(W_{i, \text{SPf}} - 7.000) + 73.32 & 0.331  & \displaystyle \frac{|w_{11}-w'_{11}|}{P} W_{i, \text{SPf}}^2 & \displaystyle 7.384 \sqrt{\sigma_{W_i, \text{SPf}}^2 + \left(\frac{\delta W_{i, \text{SPf}}}{W_{i, \text{SPf}}-7.000} \right)^2} \\ \\
	  SPs & \displaystyle f_{\text{n}, 0.5} & 63.32 W_{i, \text{SPs}} + 93.54 & -1.60\cdot 10^{-3} i_{\text{est}}+0.145 &  \delta f_{\text{n}, 0.5} & \displaystyle 63.32 \sqrt{\sigma_{W_i, \text{SPs}}^2 + \left(\delta W_{i, \text{SPs}}\right)^2} \\
      \hline
    \end{tabular}
	\end{adjustbox}
}\label{tab:formulae}
\begin{tabnote}
\end{tabnote}
\end{table*}

\begin{figure*}[]
 \begin{center}
  \includegraphics[width=0.46\textwidth]{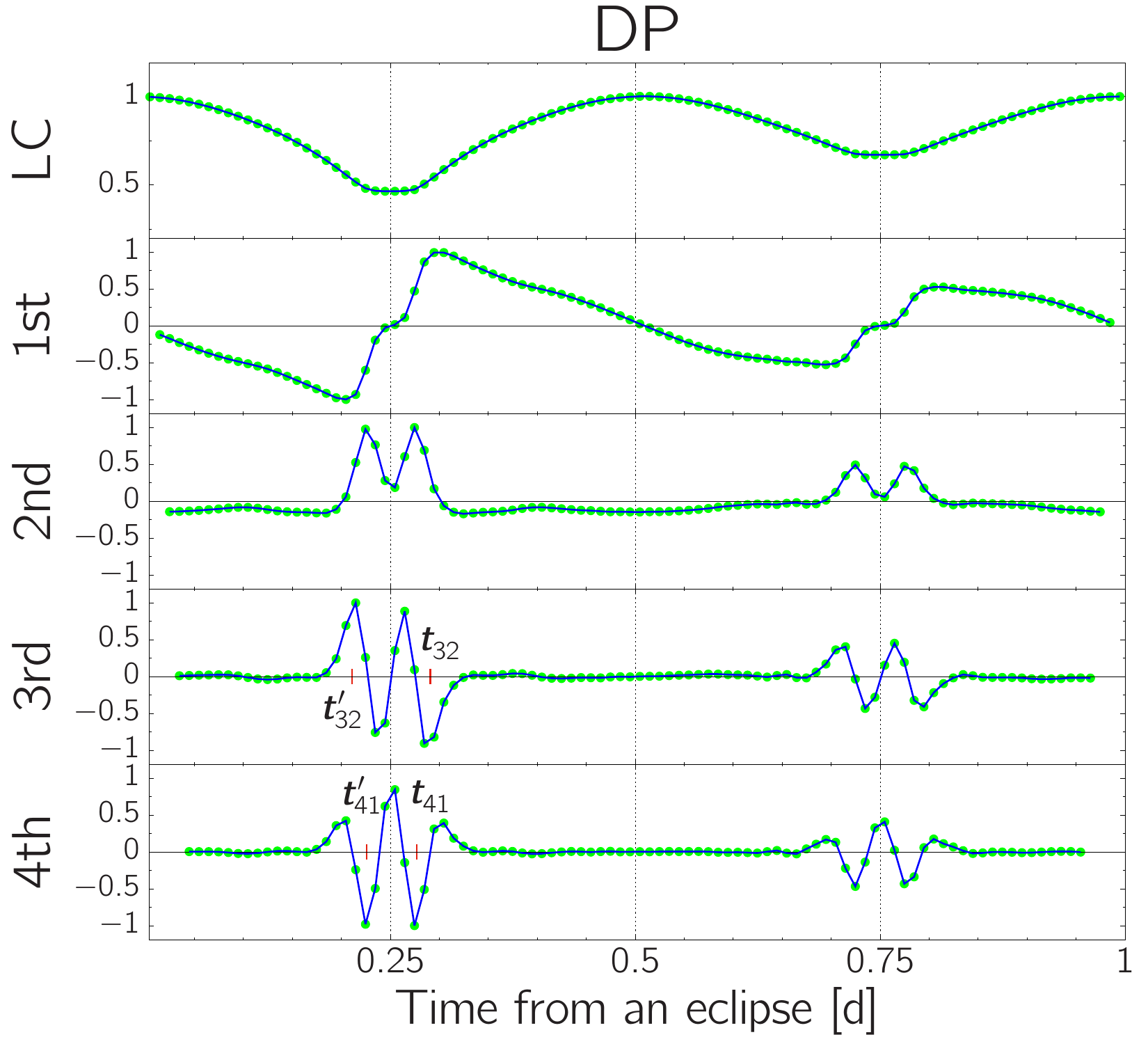}\\
  \includegraphics[width=0.46\textwidth]{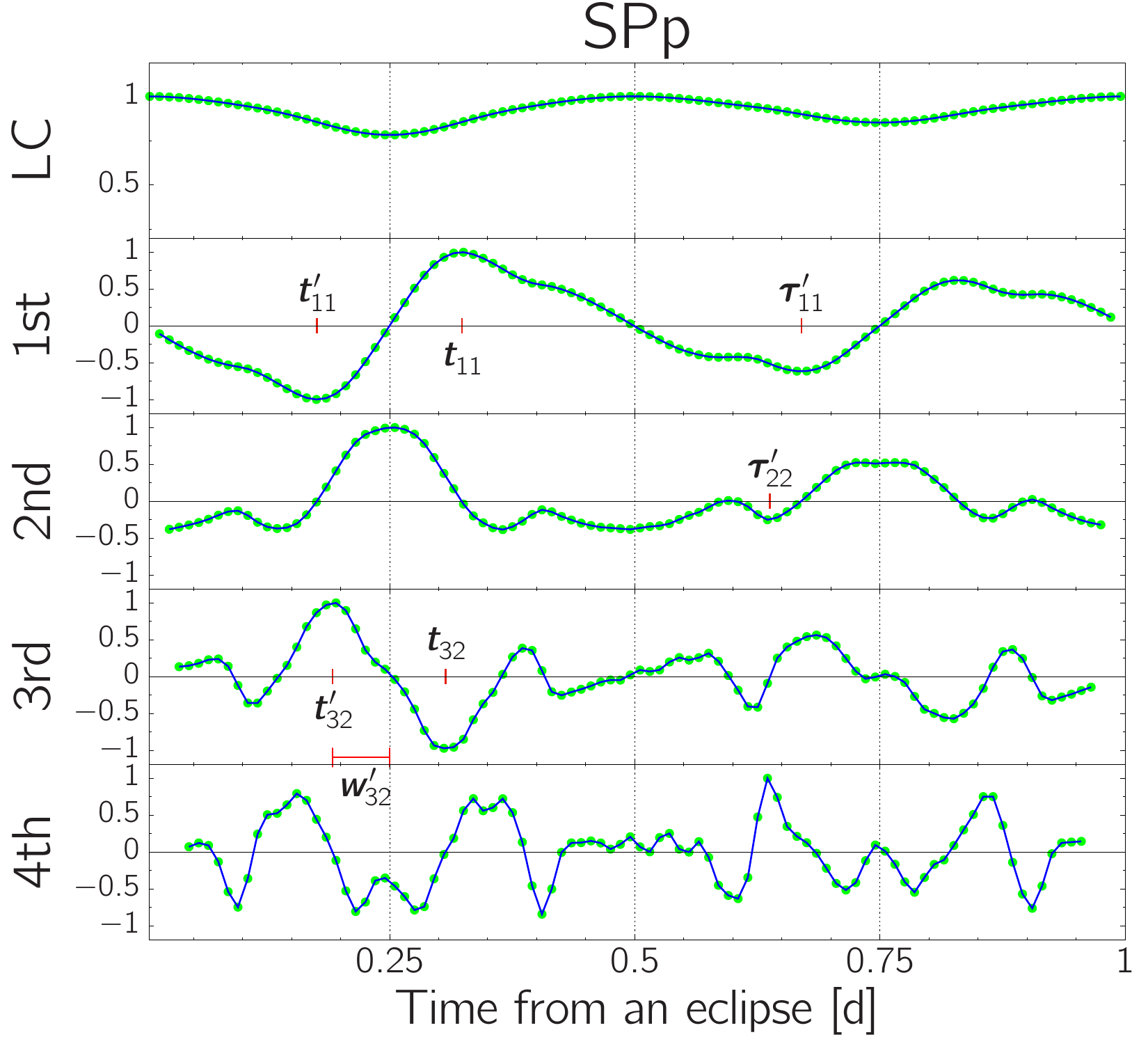}
  \includegraphics[width=0.46\textwidth]{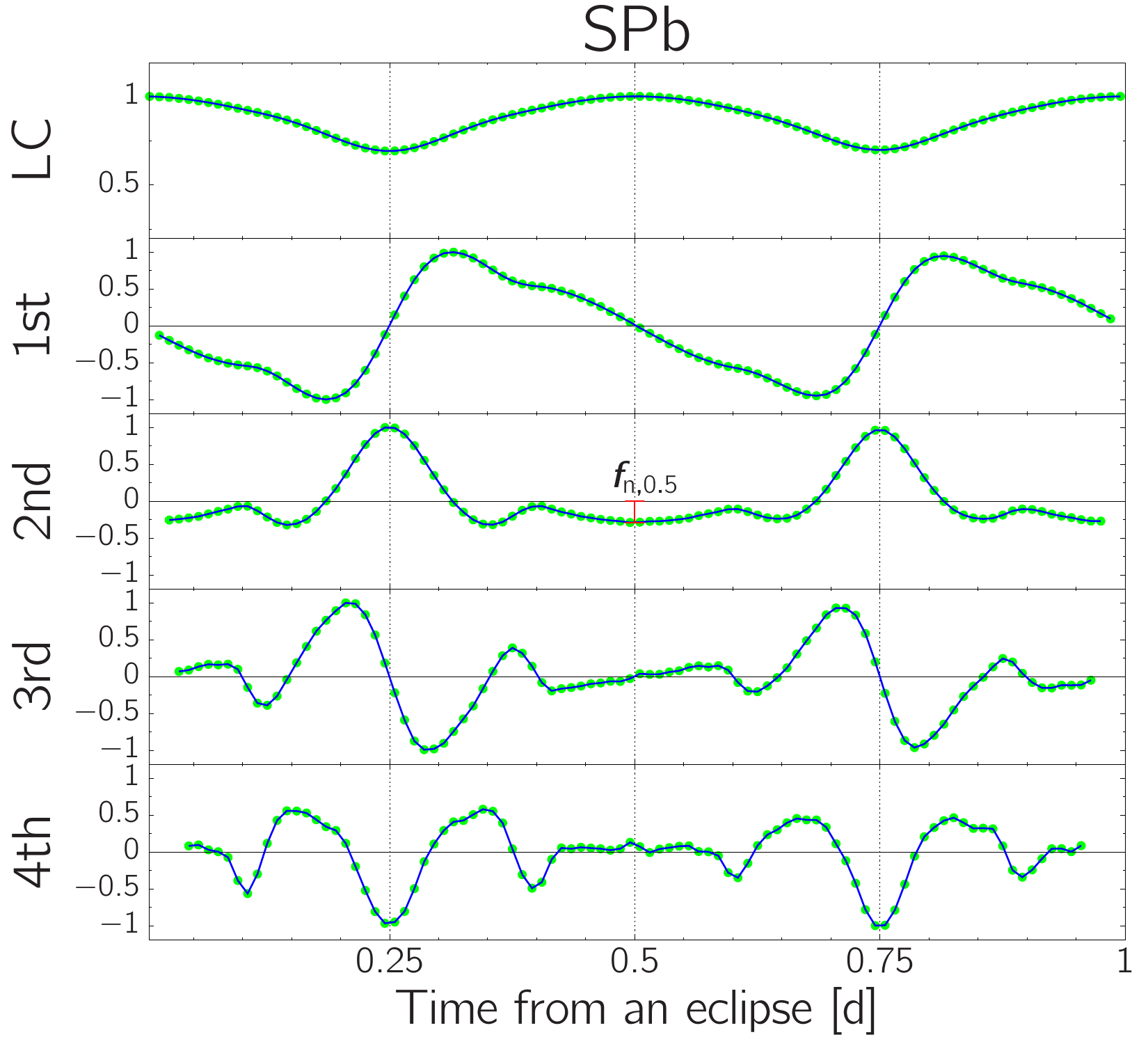}
  \includegraphics[width=0.46\textwidth]{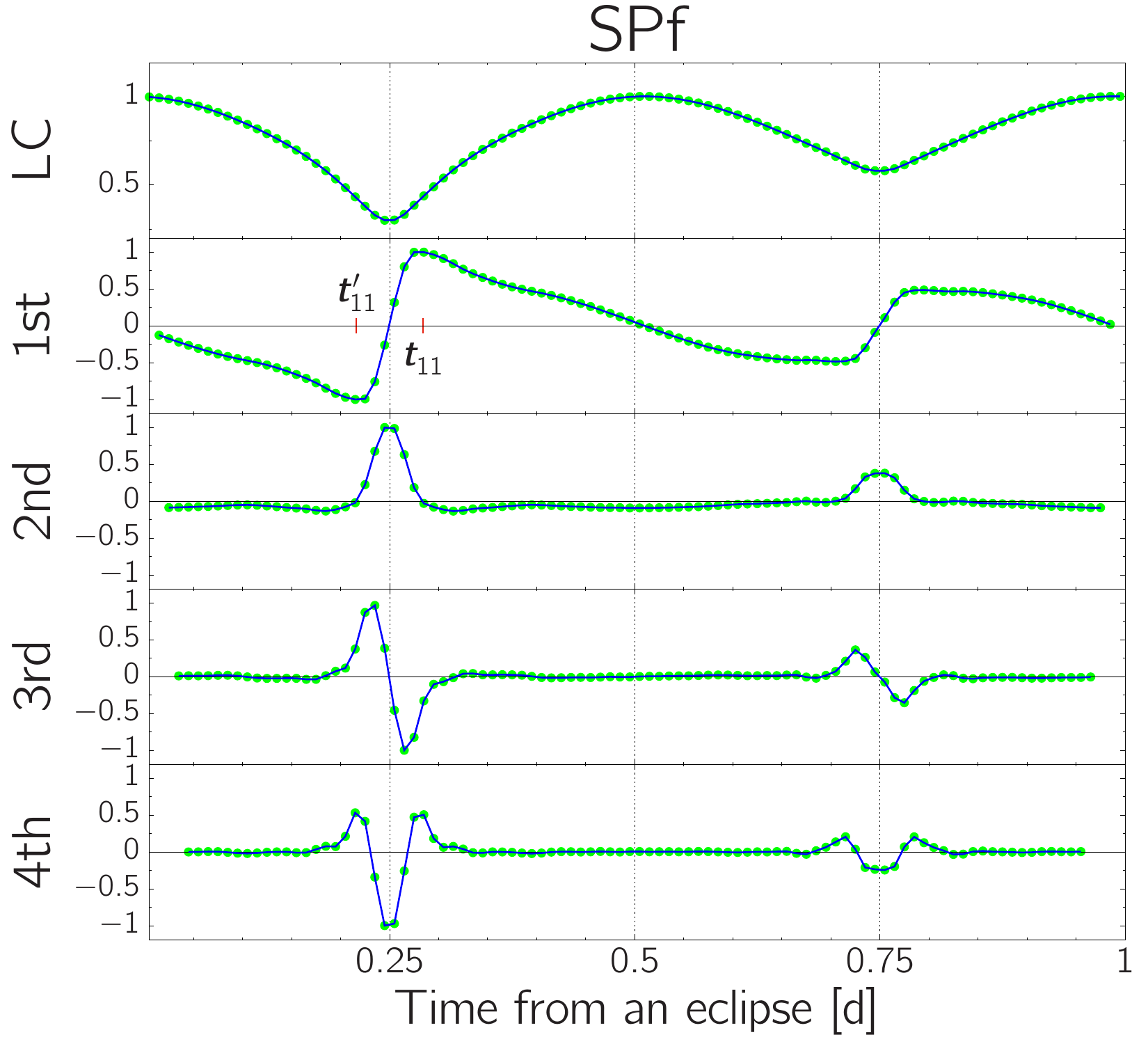}
  \includegraphics[width=0.46\textwidth]{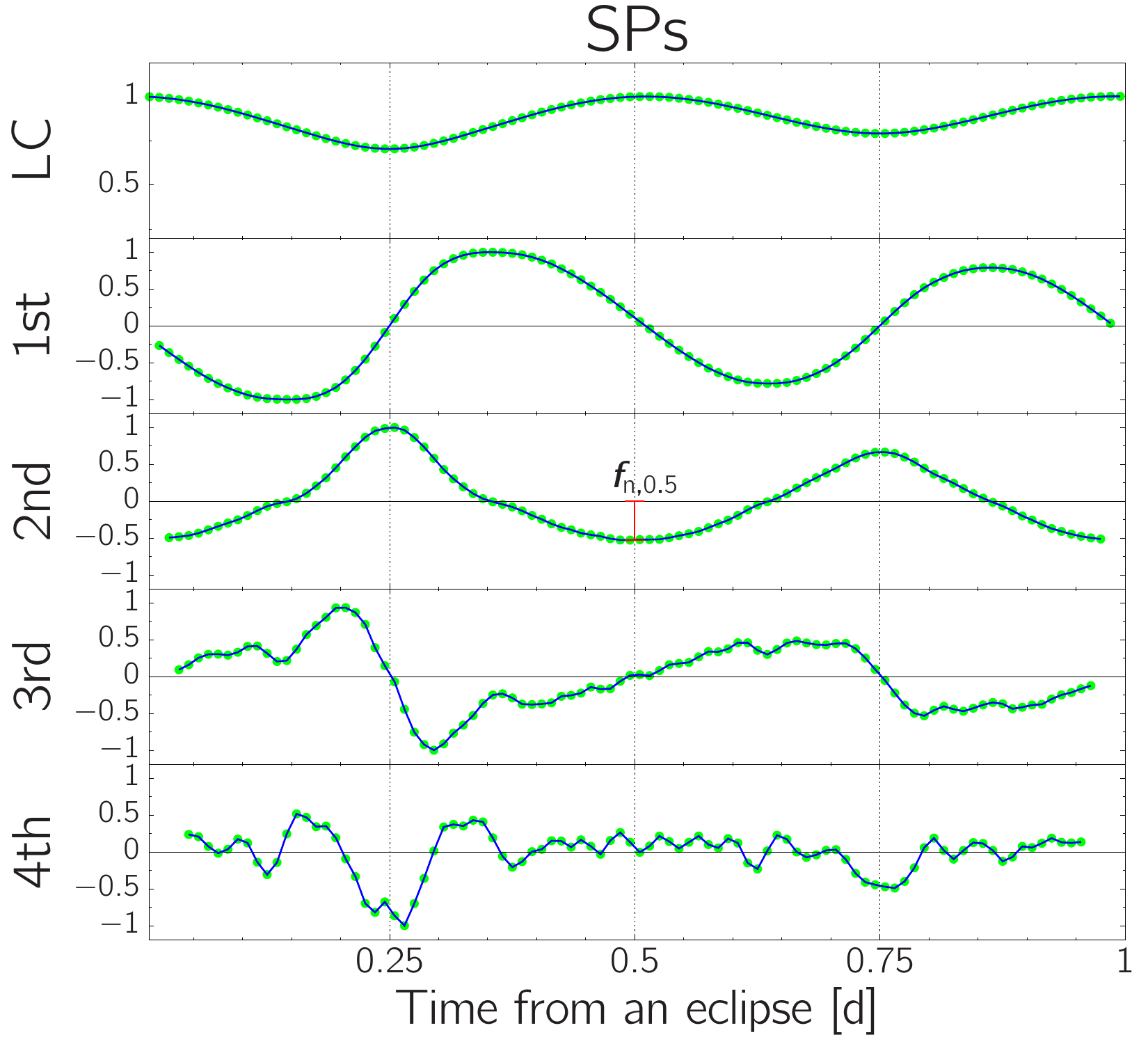}
 \end{center}

 \caption{Light curves and their first to fourth derivatives (from top to bottom) of the representative sample binaries for the five types proposed by \Kc. 
		  Each derivative is normalized by its maximum absolute value. 
		  Key times and fluxes used for estimating inclinations and their associated uncertainties are labeled. 
		 {Alt text: Five panels illustrate example light curves and their derivatives, in which key times and flux are indicated. 
		 The x-axes shows the time from an eclipse from 0 to 1. }
 \label{fig:lc-diff}}
\end{figure*}

\begin{figure*}[]
 \begin{center}
  \includegraphics[width=0.6\textwidth]{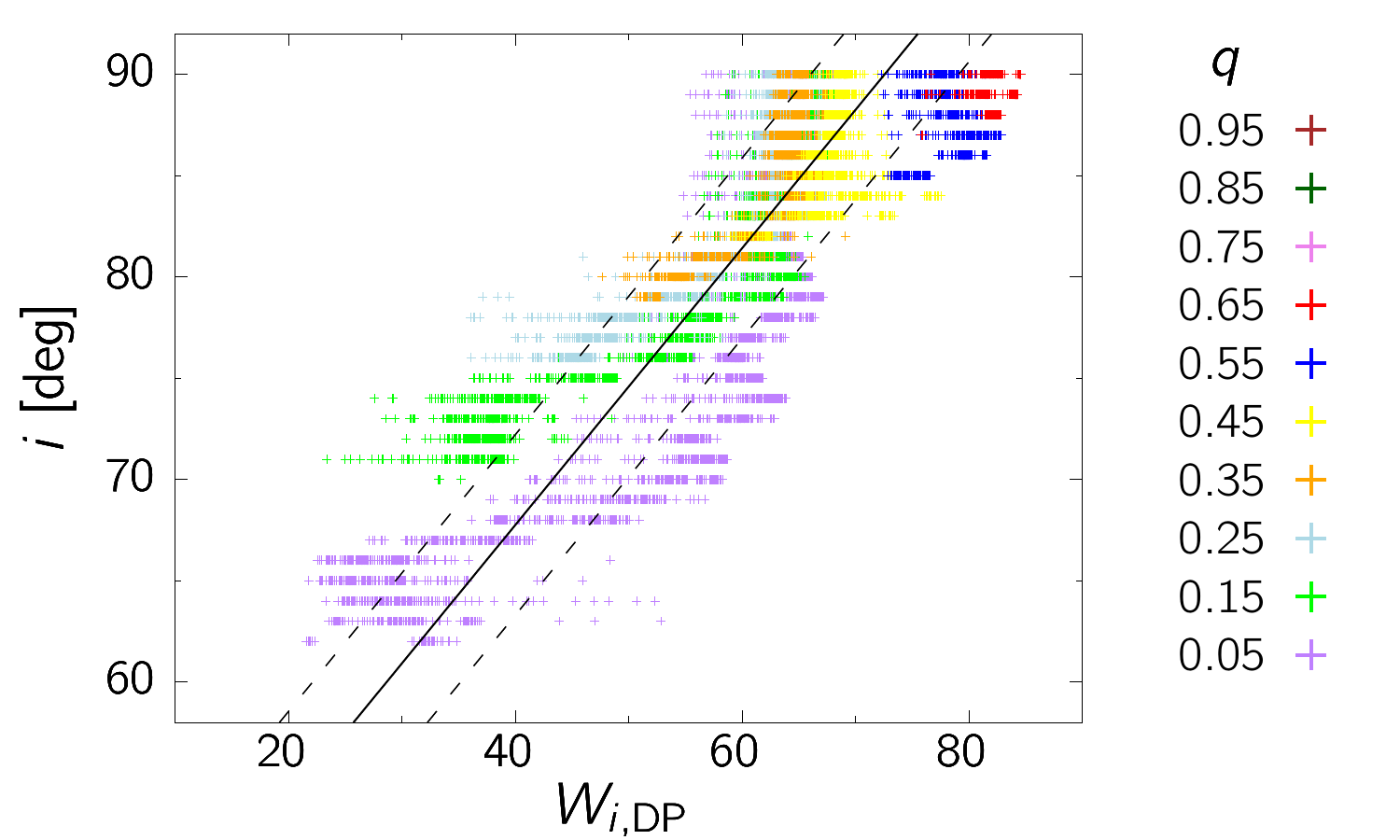}\\
  \includegraphics[width=0.48\textwidth]{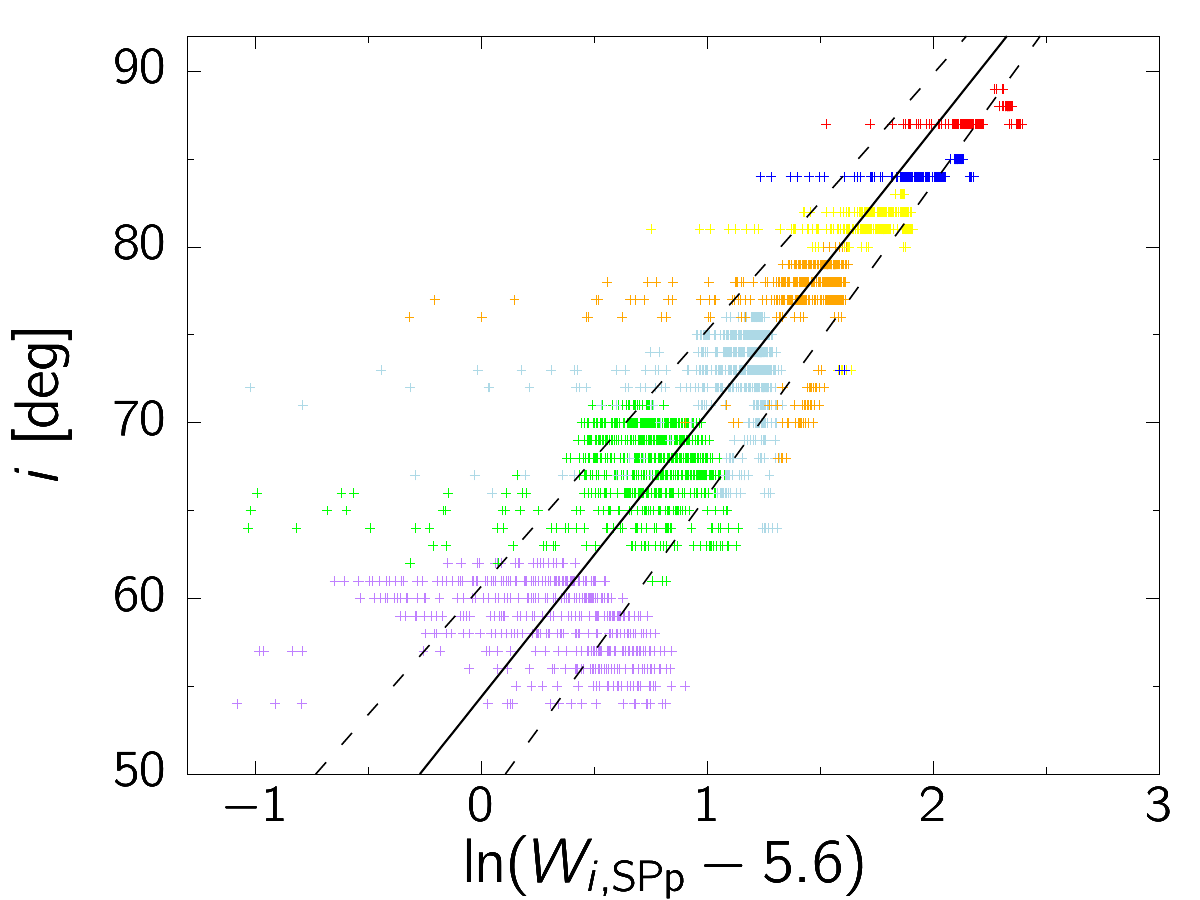}
  \includegraphics[width=0.48\textwidth]{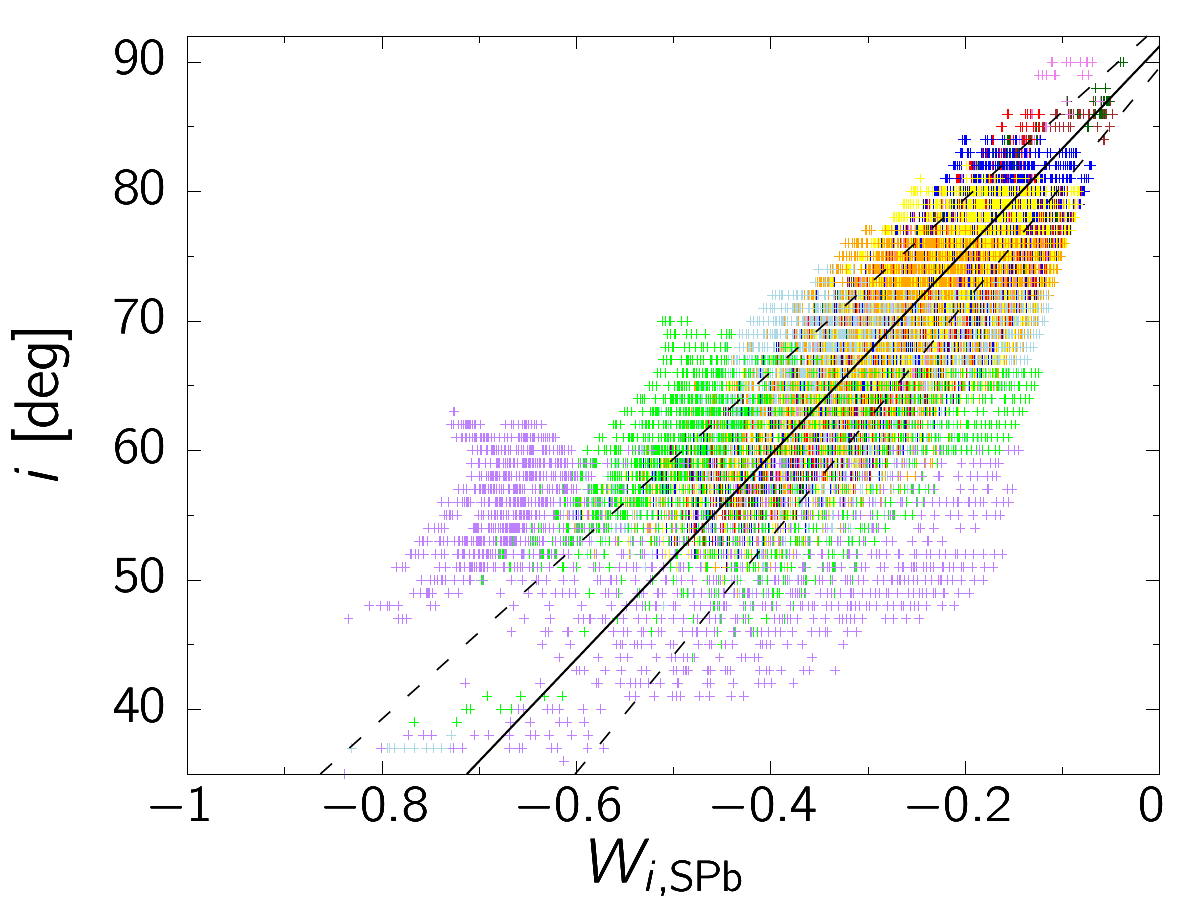}
  \includegraphics[width=0.48\textwidth]{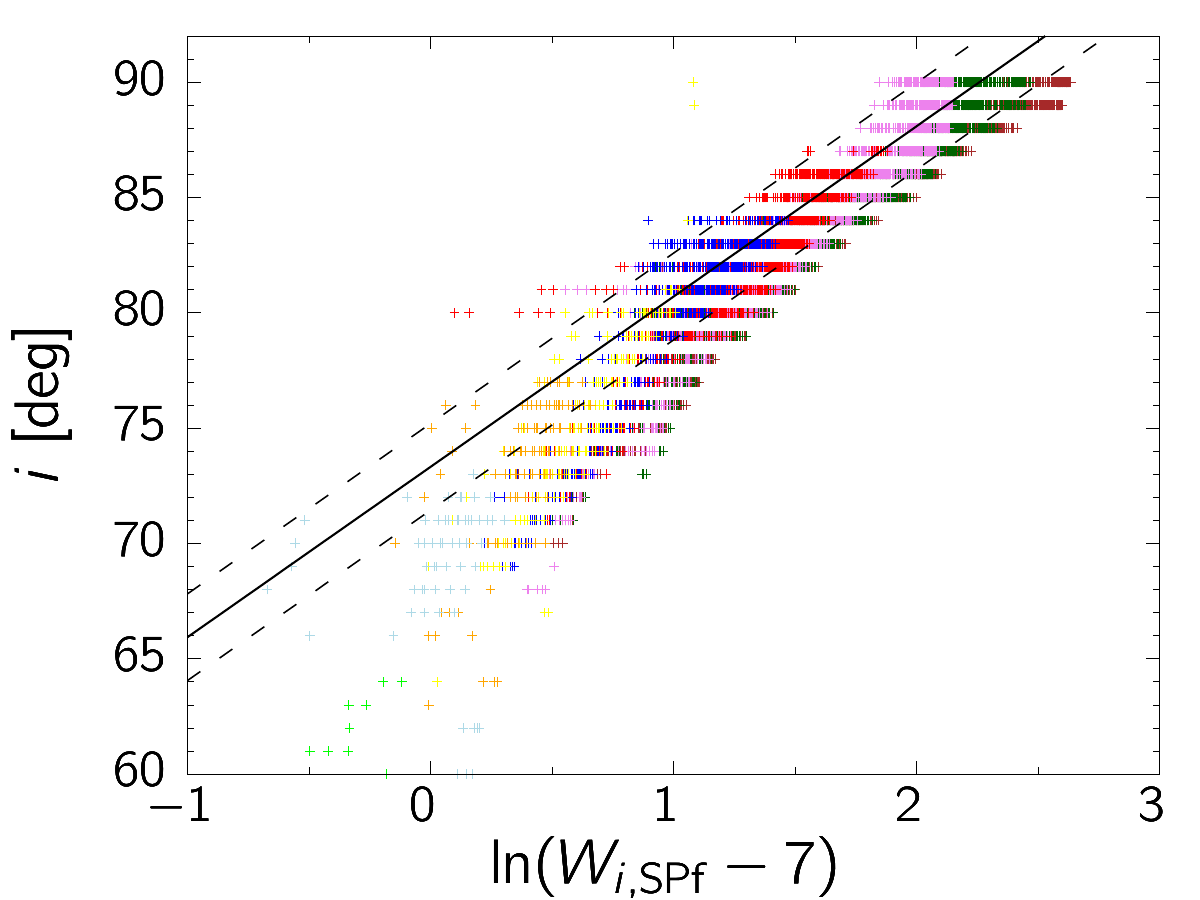}
  \includegraphics[width=0.48\textwidth]{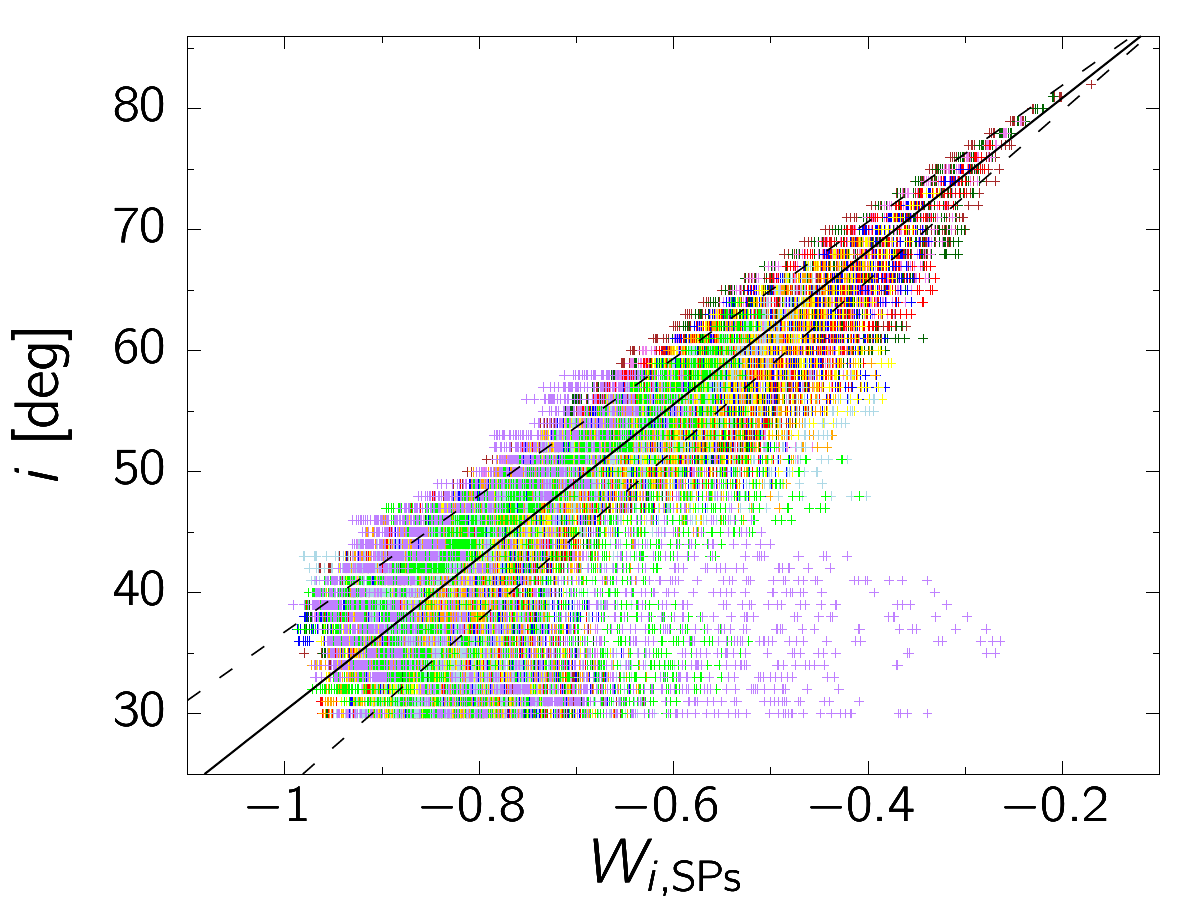}
 \end{center}
 \caption{Relationships between the $W_i$ values and orbital inclination for DP, SPp, SPb, SPf, and SPs systems. 
		  Purple, green, light blue, orange, yellow, blue, red, pink, dark green, and brown represent mass ratios of 0.05, 0.15, 0.25, 0.35, 0.45, 0.55, 0.65, 0.75, 0.85, and 0.95, respectively. 
		  The black solid lines represent the regression lines, while the upper and lower dashed lines indicate $\sigma_{W_i}$ values shown in table \ref{tab:formulae}. 
		 {Alt text: Scatter plots of the key values and inclinations. }
 \label{fig:W-i}}
\end{figure*}
\section{Method}\label{sec:method}
This paper proposes a simple method to estimate the orbital inclination angles of overcontact eclipsing binaries using the derivatives of their LCs. 
We first measured the timings of local extrema found in the derivatives of the sample LCs, together with the fluxes of characteristic extrema that are most likely to reflect the orbital inclination. 
The times of extrema were obtained by finding the zero derivatives using linear interpolation between the derivatives at consecutive phase points. 

For each of the five types (i.e., DP, SPp, SPb, SPf, and SPs), we thoroughly examined the association between the inclinations and all possible time intervals or the extremum values. 
As a result, for all five types, we identified such time intervals or extremum values that are associated with the inclinations. 
Using regression analysis, empirical formulae to estimate inclinations from these key values (referred to as $W_i$) were derived. 

Note that we selected the key values and empirical formulae in such a manner that they yield the most accurate estimates of inclination. 
This section describes the key values, empirical formulae to estimate the inclinations, and their associated uncertainties for each type.

\subsection{Inclination estimation}\label{sec:estimation}
We found that the orbital inclination is closely associated with the time intervals for DP-, SPp-, and SPf-type LCs and the depth of the local minimum at phase 0.5 for SPb- and SPs-type LCs. 
The key values ($W_i$) are summarized in column 2 of table \ref{tab:formulae}, and the times and fluxes used to compute $W_i$ are indicated in figure \ref{fig:lc-diff}. 
The time $t_{ij}$ was measured for the primary (deeper) eclipse, and $w_{ij}$ was defined as the time interval between $t_{ij}$ and the time of the primary eclipse. 
$P$ denotes the orbital period. 
We define $f_{\text{n},0.5}$ as the normalized flux at phase 0.5, which is used for SPb- and SPs-type LCs (see section \ref{sec:est_SPb} for details). 

Figure \ref{fig:W-i} shows the relationships between $W_i$ and $i$. 
Because their dependencies differ by each type, we applied linear, quadratic, or logarithmic fits to each and selected the most appropriate one. 
We performed the regression analysis by minimizing the sum of squared residuals in $W_i$ and derived empirical formulae. 
These formulae are listed in column 3 of table \ref{tab:formulae}, together with the standard deviations of the residuals along $W_i$ from the regression lines shown in figure \ref{fig:W-i} (i.e., $\sigma_{W_i}$ in column 4). 
If the standard deviation varies significantly with the inclination, we apply a linear fit and derive the standard deviation as a function of the inclination. 

Our estimation method involves several considerations that warrant attention. 
The following describes important aspects to be taken into account during estimation for each type.

\subsubsection{DP}
DP-type LCs are characterized by the appearance of a double peak around the time of an eclipse in the second derivative of a LC. 
The key value $W_{i,\text{DP}}$ should be computed for the eclipse that exhibits a double peak. 
If such a double peak is observed around both eclipses, we use the mean of both the $W_{i,\text{DP}}$ values and also calculate the uncertainty described in section \ref{sec:uncertainty} from both the $W_{i,\text{DP}}$ values.

\subsubsection{SPp}
SPp-type LCs display a peak in the fourth derivative at an eclipse. 
When this feature is found around the primary eclipse or around both eclipses, we compute the value of $W_{i,\text{SPp}}$ using the time interval around the primary eclipse. 
If the feature appears solely near the secondary eclipse, $W_{i,\text{SPp}}$ is generally computed from the time interval around the secondary eclipse. 
However, when the times $\tau'_{11}$ and $\tau'_{22}$ (see figure \ref{fig:lc-diff}) cannot be determined or do not satisfy the condition  
\begin{equation}
	\tau'_{11} > \tau'_{22}, 
\end{equation}
the interval around the primary eclipse should be used instead to ensure a more reliable estimate. 

Figure \ref{fig:W-i} shows that the inclination increases with increasing mass ratio. 
This indicates that, in SPp systems, the inclination is also closely associated with the mass ratio. 

As mentioned in \Kc, the peak in the fourth derivative is prone to be misidentified. 
In such cases, the estimated inclination may deviate from the true value.

\begin{figure}[]
 \begin{center}
  \includegraphics[width=0.48\textwidth]{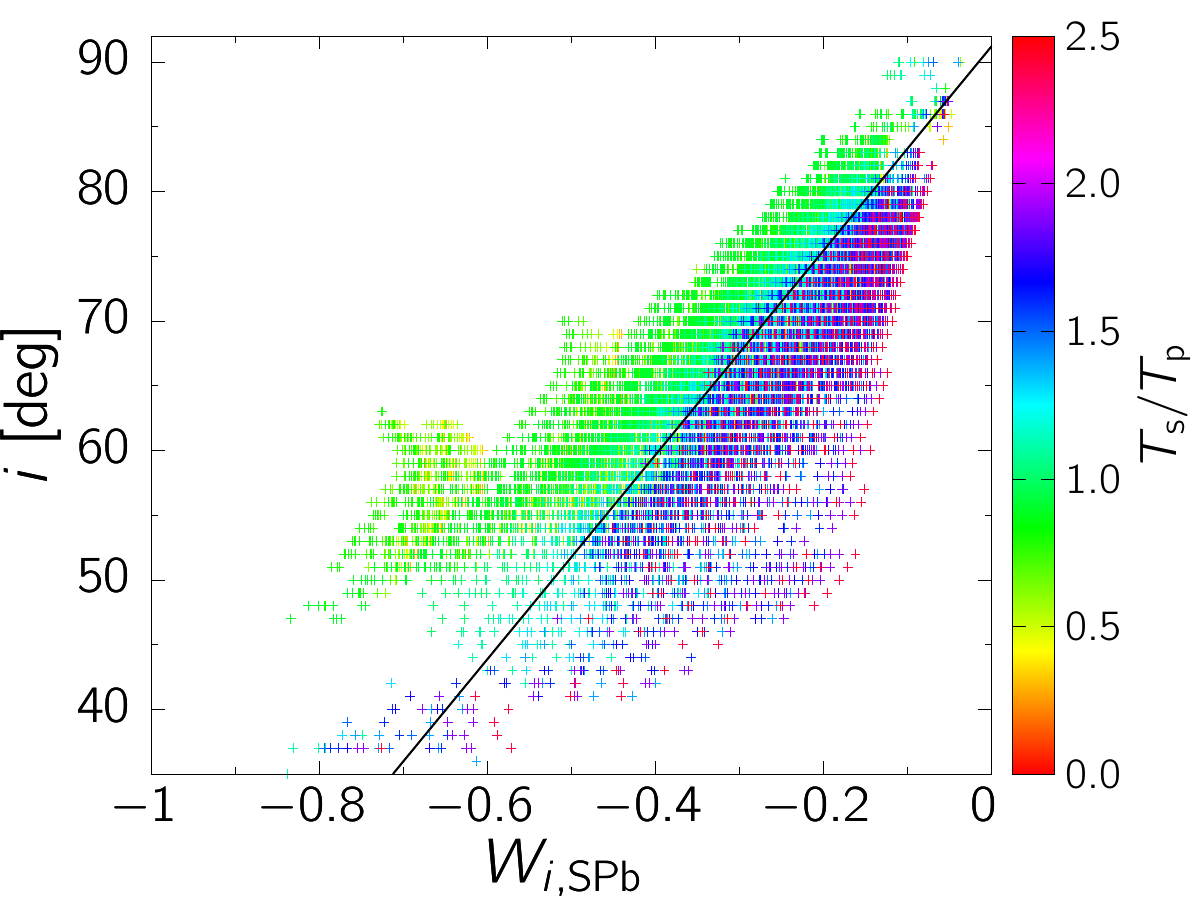}
 \end{center}
 \caption{Scatter plot of $W_{i,\text{SPb}}$ and $i$, showing temperature ratio differences as a color gradient. 
			{Alt text: The x-axis shows $W_{i,\text{SPb}}$ value ranging from -1 to 0, and the y-axis shows the inclination from $35^\tcdegree$ to $90^\tcdegree$. }
 \label{fig:W-i-SPb-Trat}}
\end{figure}
\subsubsection{SPb}\label{sec:est_SPb}
The shapes of an SPb-type LC and its derivatives are often similar to those of SPp-type LCs, but they exhibit no peak in the fourth derivative around an eclipse. 
The inclination of SPb systems is closely related to the depth at phase 0.5 in the second derivative. 
This depth (normalized flux) at phase 0.5 ($f_{\text{n},0.5}$) was obtained using the following quadratic fit: 
\begin{equation}\label{eq:qfit}
	f(x) = c_2 (x+c_1)^2 + c_0, 
\end{equation}
where $c_2$, $c_1$, and $c_0$ are regression coefficients. 
These coefficients were optimized using the Levenberg--Marquardt method. 
Among the obtained coefficients, $c_0$ and its standard error from the optimization were taken as the value of $f_{\text{n},0.5}$ and its corresponding uncertainty $\delta f_{\text{n},0.5}$. 

Figure \ref{fig:W-i} indicates that systems with smaller actual orbital inclinations tend to have smaller mass ratios, and their distributions are more widely dispersed. 
This indicates that SPb systems with smaller mass ratios tend to have larger estimation uncertainties. 

The primary factor contributing to the dispersion at an inclination is the difference in the temperature ratio between two component stars (i.e., $T_\text{s}/T_\text{p}$). 
Figure \ref{fig:W-i-SPb-Trat} shows the relationship between $W_{i,\text{SPb}}$ and $i$, with differences in the temperature ratio represented by a color gradient. 
As seen in the figure, the $W_{i,\text{SPb}}$ values generally tend to increase with increasing temperature ratio for a given inclination.

\subsubsection{SPf}
SPf-type LCs exhibit flat shapes around phase 0.5 in the second to fourth derivatives. 
The value $W_{i,\text{SPf}}$ is calculated on the basis of the time interval between the local minimum and maximum in the first derivative ($t'_{11}$ and $t_{11}$). 
This calculation must use times around the primary eclipse. 

For SPf systems, the distribution in the $W_i$--$i$ plot exhibits a noticeable dependence on whether the temperature ratio of the components is close to unity or significantly different. 
Observationally, many overcontact binaries have component stars with almost identical temperatures. 
Taking this into account, the regression line was derived using a weighting scheme that assigned greater weight to systems with nearly equal temperatures.

\begin{figure}[]
 \begin{center}
  \includegraphics[width=0.48\textwidth]{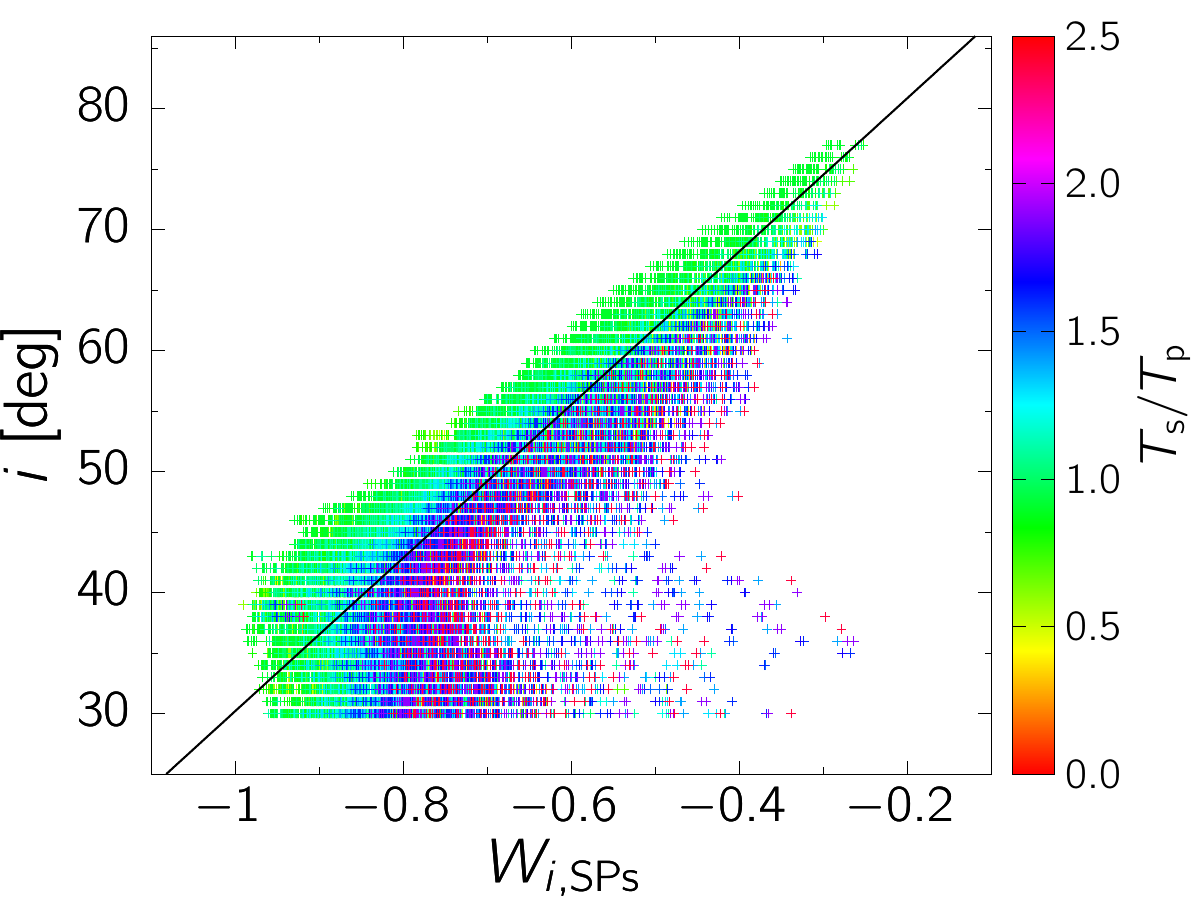}
 \end{center}
 \caption{Scatter plot of $W_{i,\text{SPs}}$ and $i$, showing temperature ratio differences as a color gradient. 
			{Alt text: The x-axis shows $W_{i,\text{SPs}}$ value ranging from -1 to 0, and the y-axis shows the inclination from $25^\tcdegree$ to $85^\tcdegree$. }
 \label{fig:W-i-SPs-Trat}}
\end{figure}
\subsubsection{SPs}
SPs-type LCs display smooth curves in their second derivatives. 
The depth at phase 0.5 in the second derivative of a LC is closely associated with the inclination, as is the case for SPb-type LCs. 
Accordingly, $W_{i,\text{SPs}}$ is defined in the same manner as $W_{i,\text{SPb}}$, that is, $W_{i,\text{SPs}} = f_{\text{n},0.5}$.  

A relatively large dispersion in the figure is primarily due to differences in the temperature ratio of the two components, similar to SPb-type LCs. 
Figure \ref{fig:W-i-SPs-Trat} demonstrates that, at a given inclination, $W_{i,\text{SPs}}$ tends to increase for systems with higher temperature ratios. 

In the scatter plot for SPs systems, although the inclination is closely correlated with the $W_{i, \text{SPs}}$ above $i\sim 38 \tcdegree$, no significant association is found below this inclination. 
Most SPs systems with $i<38\tcdegree$ also tend to have low mass ratios (see figure \ref{fig:W-i}), resulting in LCs with extremely shallow eclipses. 
Accordingly, their derivatives were noisy, making it difficult to determine accurate $W_{i,\text{SPs}}$ values. 
In such situations, quadratic fits were unreliable. 
Taking this into account, we excluded the SPs systems with $i<38\tcdegree$ from the regression analysis.

\subsection{Uncertainty}\label{sec:uncertainty}
We estimated the uncertainty of the derived inclination by taking into account both the measurement uncertainty in the timings of local extrema ($\delta W_i$) and the uncertainty in the empirical formulae ($\sigma_{W_i}$). 
The uncertainty in the orbital period was assumed to be sufficiently small to be negligible. 

The empirical formulae in table \ref{tab:formulae} exhibit that the inclination is a function of $W_i$. 
Accordingly, the uncertainty in the inclination is derived by 
\begin{equation}
	\delta i =  \left|\frac{di}{dW_i} \right| \sqrt{\sigma_{W_i}^2 + \left(\delta W_i \right)^2} 
\end{equation}
for the linear $i = aW_i + b$ regression model and 
\begin{equation}
	\delta i =  \sqrt{\left(\frac{di}{d(\ln(W_i+b))}\sigma_{W_i} \right)^2 + \left(\frac{di}{dW_i} \delta W_i \right)^2} 
\end{equation}
for the logarithmic $i=a\ln(W_i+b)+c$ regression model. 

For SPf-type LCs, $\delta W_i$ can be evaluated in the same manner as in \citet{Kouzuma2023-ApJ}. 
Specifically, assuming that the two local extrema in the derivatives of a LC are perfectly symmetric (i.e., $w_{11}=w'_{11}$), the standard uncertainty of $W_i$ is given by 
\begin{equation}\label{eq:deltaW}
	\delta W_{i,\text{SPf}} = \frac{|w_{11}-w'_{11}|}{P} W_{i,\text{SPf}}^2. 
\end{equation}

For DP- and SPp-type LCs, $W_i$ values are taken as the mean of two time intervals. 
When the two time intervals ($d'$ and $d$) are assumed to be equal, we derive the standard uncertainty as 
\begin{equation}
	\delta W_{i} = \frac{P}{2\sqrt{2}} \left(\frac{1}{d'^4}+\frac{1}{d^4}\right)^\frac{1}{2} |d-d'|. 
\end{equation}
In this situation, $\delta W_i$ values for DP and SPp types are calculated using  
\begin{align}
	d'_\text{DP} &= t'_{41}-t'_{32} = w'_{32}-w'_{41}, \\
	d_\text{DP} &= t_{32}-t_{41} = w_{32}-w_{41},
\end{align}
and 
\begin{align}
	d'_\text{SPp} &= t_{32}-t'_{11} = w_{32}+w'_{11}, \\
	d_\text{SPp} &= t_{11}-t'_{32} = w_{11}+w'_{32},
\end{align}
respectively. 

For SPb- and SPs-type LCs, $W_i$ values are obtained from $f_{\text{n},0.5}$. 
As described in section \ref{sec:est_SPb}, $f_{\text{n},0.5}$ values are derived from quadratic fits using the Levenberg--Marquardt method. 
Accordingly, their uncertainties ($\delta f_{\text{n},0.5}$) were taken as the standard errors of the coefficient $c_0$ in equation (\ref{eq:qfit}) obtained from the quadratic fits. 

In addition to the derived $\delta W_i$ values, using the $\sigma_{W_i}$ values in table \ref{tab:formulae}, we derive the uncertainty of the estimated inclination. 
An expanded uncertainty is obtained through the multiplication of these values by an appropriate coverage factor.

\begin{table}
  \tbl{Numbers of A-subclass, W-subclass, and total systems for each classified LC type. 
	   The numbers in parentheses indicate the size of the spectroscopic sample. 
  }{%
  \begin{tabular}{rrrrrr}
      \hline
      \hline
      LC type 	  & DP & SPp & SPb & SPf & SPs \\
	  A-subclass  & 143 (33) & 13 (4) & 32 (17) & 10 (3) &  7 (4)    \\
	  W-subclass  & 123 (38) & 15 (8) & 38 (20) & 15 (2) &  10 (5)   \\
	  Total		  & 266 (71) &	28 (12) & 70 (37) & 25 (5) & 17 (9)  \\
      \hline
    \end{tabular}}\label{tab:numbers}
\begin{tabnote}
\end{tabnote}
\end{table}

\section{Application to real binary data}\label{sec:application}
To evaluate the effectiveness of the proposed method, we applied it to observed overcontact binary systems. 
We used the sample from the catalog compiled by \citet{Latkovic2021-ApJS}. 
This catalog lists binary parameters obtained in previous studies either through photometric alone or through combined photometric and spectroscopic analyses. 
Hereafter, we refer to the former as the photometric sample and the latter as the spectroscopic sample. 

Using the python package Lightkurve \citep{Lightkurve2018-code}, we extracted LCs from the TESS and Kepler archival data and phase-folded them using the orbital periods determined via periodogram analysis. 
When two or more LCs were available for a binary, the better-quality one was selected in such a manner that its derivatives were less noisy and smoother than the other one. 
Apparent outliers in the LCs were removed in advance. 

Our sample consists of 153 spectroscopic and 410 photometric sample systems with extracted LCs. 
After computing the LC derivatives up to the fourth order, we classified the LCs into the five types on the basis of \Kc. 
Excluding systems whose LCs could not be classified, our final dataset comprised 134 spectroscopic and 272 photometric sample systems.
Table \ref{tab:numbers} summarizes the number of each type, together with their A- or W-subclass classification proposed by \citet{Binnendijk1970-VA}. 
We applied the method described in section \ref{sec:method} to the selected LCs of the sample systems. 
For each of the five types, we then derived inclination estimates and their associated uncertainties.

\begin{figure*}[]
 \begin{center}
  \includegraphics[width=0.48\textwidth]{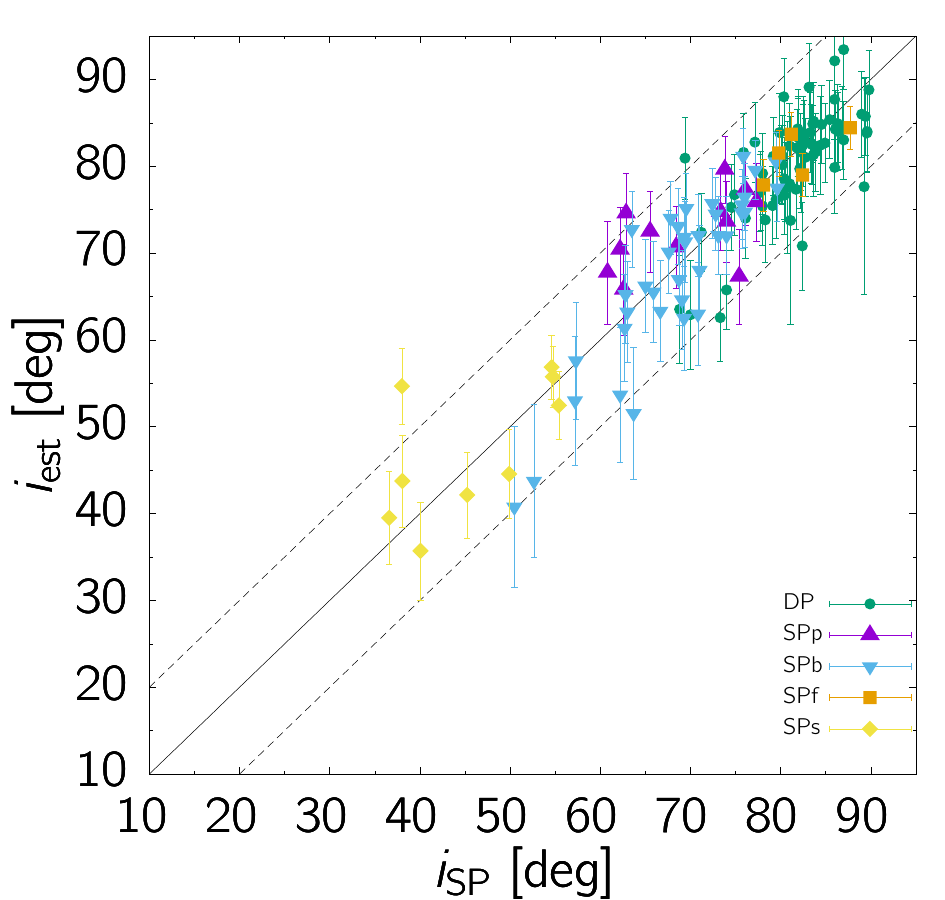}
  \includegraphics[width=0.48\textwidth]{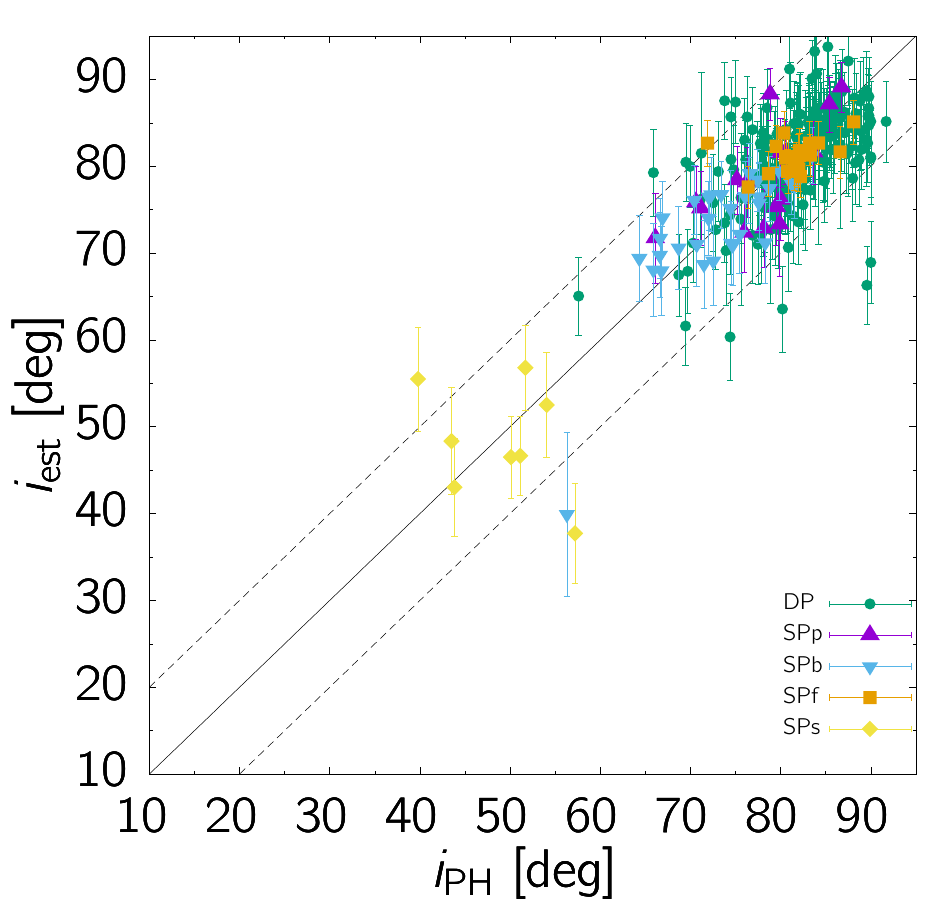}
 \end{center}
 \caption{Comparison between the orbital inclinations estimated by the proposed method ($i_\text{est}$) and those reported in the literature ($i_\text{SP}$ for the spectroscopic sample and $i_\text{PH}$ for the photometric sample). 
		  The solid lines indicate $i_\text{est}=i_\text{SP}$ and $i_\text{est}=i_\text{PH}$, and the dashed lines represent offsets of $\pm 10^\tcdegree$ from the solid lines. 
		  Green circles, purple triangles, blue inverted triangles, orange squares, and yellow diamonds correspond to DP, SPp, SPb, SPf, and SPs systems, respectively. 
		 {Alt text: Scatter plots comparing the estimated and reported inclinations. Both x- and y-axes show inclinations ranging from $10^\tcdegree$ to $90^\tcdegree$. }
 \label{fig:comparison}}
\end{figure*}

\begin{figure*}[]
 \begin{center}
  \includegraphics[width=0.43\textwidth]{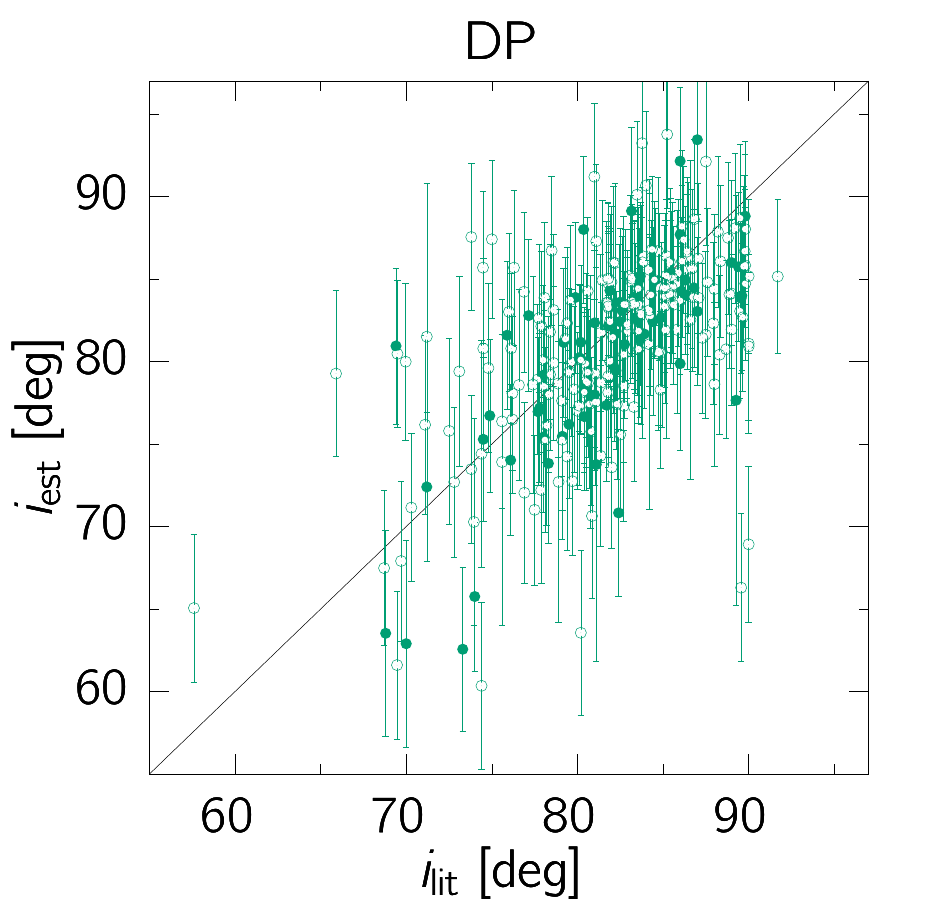}\\
  \includegraphics[width=0.43\textwidth]{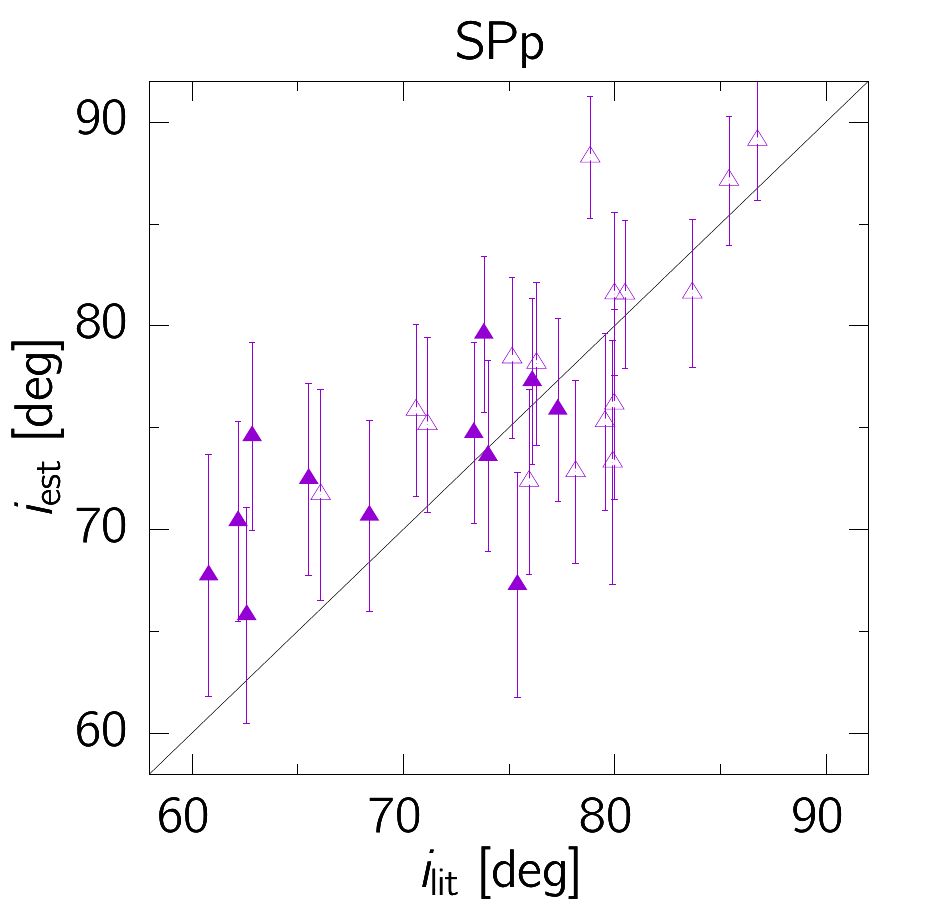}
  \includegraphics[width=0.43\textwidth]{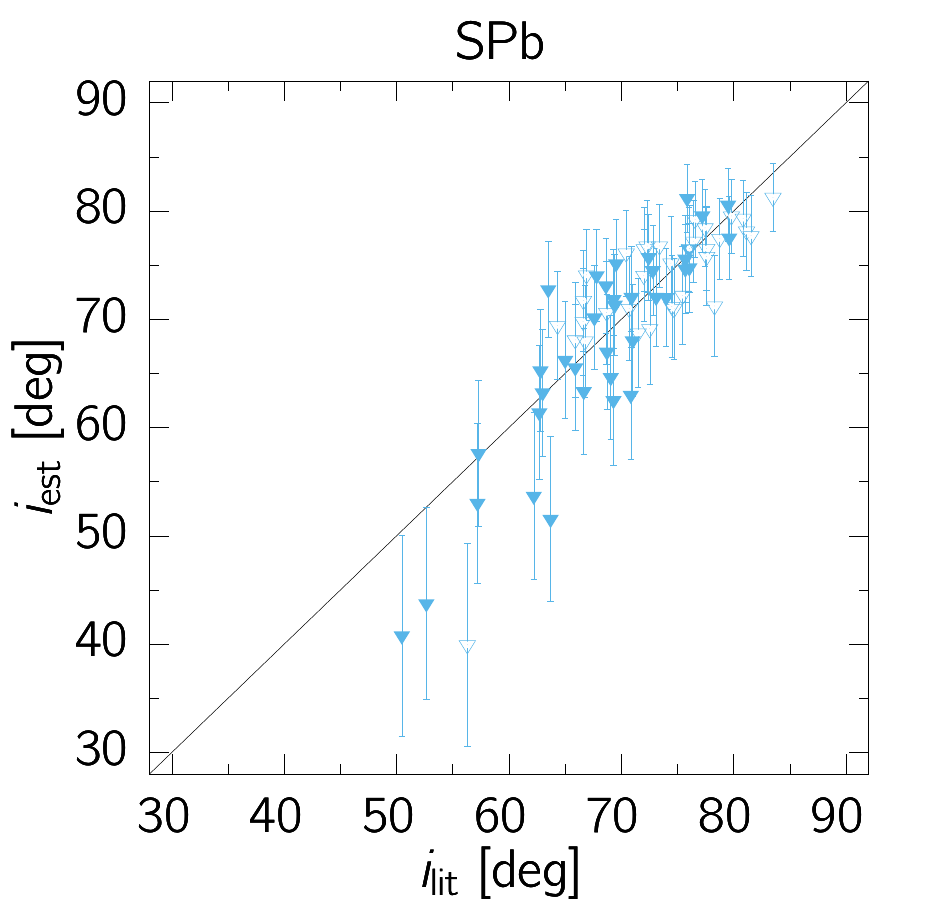}
  \includegraphics[width=0.43\textwidth]{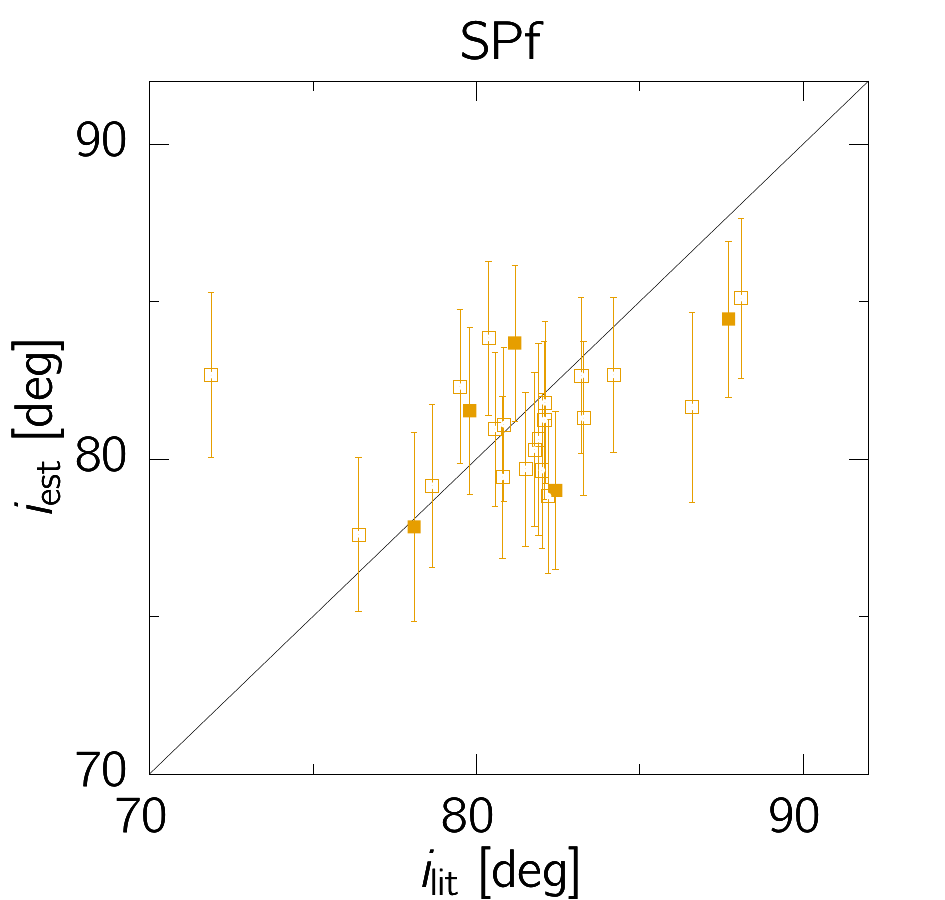}
  \includegraphics[width=0.43\textwidth]{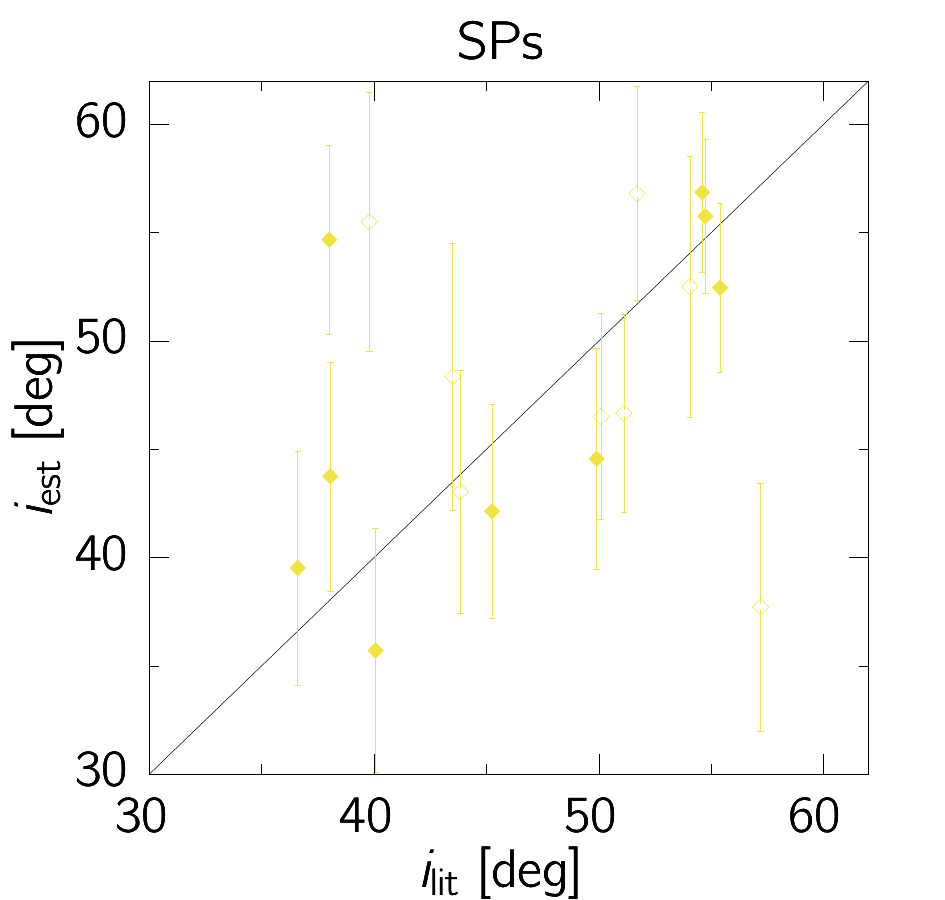}
 \end{center}
 \caption{Comparison between the orbital inclinations estimated by the proposed method ($i_\text{est}$) and those reported in the literature ($i_\text{lit}$). 
		  Filled and open symbols correspond to the spectroscopic and photometric samples, respectively. 
		 {Alt text: Scatter plots comparing the estimated and reported inclinations for each of the five types: top (DP, $55^\tcdegree$-$90^\tcdegree$), middle-left (SPp, $60^\tcdegree$-$90^\tcdegree$), middle-right (SPb, $30^\tcdegree$-$90^\tcdegree$), bottom-left (SPf, $70^\tcdegree$-$90^\tcdegree$), and bottom-right ($30^\tcdegree$-$60^\tcdegree$) panels. 
		 }
 \label{fig:comparison_each}}
\end{figure*}

\begin{table}
  \tbl{Fraction of systems with estimated inclinations consistent with literature within uncertainties. 
	   The decimals indicate percentages, and the numbers in parentheses indicate the number of systems. 
  }{%
  \begin{tabular}{rrrrrr}
      \hline
      \hline
      LC type 			   & DP & SPp & SPb & SPf & SPs \\
	  Spectroscopic sample & 80.3 (57)  & 50.0 (6)  & 70.3 (26) & 40.0 (2) & 66.7 (6)    \\
	  Photometric sample   & 71.8 (140) & 68.8 (11) & 72.7 (24) & 70.0 (14) & 62.5 (5)    \\
	  All sample		   & 74.1 (197) & 60.7 (17) & 71.4 (50) & 64.0 (16) & 64.7 (11)  \\
      \hline
    \end{tabular}}\label{tab:effectiveness}
\begin{tabnote}
\end{tabnote}
\end{table}
\section{Results and Discussion}\label{sec:result}
\subsection{Effectiveness of our method}
Figure \ref{fig:comparison} shows a comparison between our inclination estimates ($i_\text{est}$) and corresponding literature values for the spectroscopic ($i_\text{SP}$) and photometric ($i_\text{PH}$) samples. 
No significant differences seem to be observed between the spectroscopic and photometric samples. 
In both samples, approximately 95\% of the estimated inclinations agree with the literature values within $\pm 10\tcdegree$. 

Figure \ref{fig:comparison_each} separates the data used for comparing the estimated and literature ($i_\text{lit}$) inclinations by system type. 
Table \ref{tab:effectiveness} summarizes, for each type, the percentage of estimated inclinations that agree with literature values within the estimated uncertainties. 
For all five types, 61-74\% of the systems show agreement between the estimated and literature values within the estimated uncertainty, which is close to the 68\% level associated with the standard uncertainty. 
The uncertainties of inclinations reported in the literature are typically extremely small, with almost all being well below 1\% of the inclination value. 
While several authors have pointed out that such uncertainties are unrealistically small \citep{Maceroni1997-PASP,Pavlovski2009-MNRAS,Southworth2011-MNRAS}, some studies have evaluated the uncertainties and found that they can reach a few degrees \citep{Maceroni1997-PASP,Liu2021-PASP}. 
Taking this into account, if such uncertainties are considered in addition to our estimated uncertainties, the fractions in table \ref{tab:effectiveness} are likely to increase slightly. 

In table \ref{tab:effectiveness}, the fractions of DP type are slightly higher than those of other types, which is likely due to a selection effect. 
As seen in figure \ref{fig:W-i}, the distribution between $i=80^\tcdegree$ and $85^\tcdegree$ exhibits relatively small dispersion and thus a relatively small standard deviation. 
Half of the DP systems have inclinations within this range. 
For DP systems, although the uncertainty based on the standard deviation slightly varies with the inclination, we assume a constant uncertainty. 
Since the dispersion of inclinations is relatively small within the range $80^\tcdegree$--$85^\tcdegree$, this assumption tends to overestimate the uncertainty, which likely explains the slightly higher fraction of DP type shown in table \ref{tab:effectiveness}. 
If we could collect a sample without bias in the possible orbital inclinations, this fraction is expected to decrease slightly. 

In summary, the estimated inclinations generally agree with the corresponding literature values. 
The fractions of agreement are close to 68\%, though some variations exist among the different types. 
The observed differences are likely due to differences in sample sizes, selection effects, and the unclear true uncertainties of the literature inclinations. 
Consequently, in practical terms, our proposed method is expected to provide reasonably reliable inclination estimates and their associated uncertainties.

\subsection{Possible factors affecting the accuracy of the proposed method}
The presence of light from the third body (third light) decreases the amplitude of the variations in the LC of the eclipsing binary when the LC is expressed in magnitude. 
The third light can be assumed to be nearly constant in most cases. 
Let the intrinsic LC in flux be $F(t)$ and the flux of the third light be a constant $F_3$. 
The observed LC then becomes 
\begin{equation}
	F_\textnormal{obs}(t) = F(t) + F_3. 
\end{equation}
Taking the derivative gives 
\begin{equation}
	\frac{dF_\textnormal{obs}(t)}{dt} = \frac{dF(t)}{dt}, 
\end{equation}
indicating that the derivative is unchanged by the addition of a constant third light. 
In our analysis, the derivatives are normalized by the maximum absolute value of the derivative, so the normalized derivatives remain unchanged. 
In this situation, the third light does not affect the derivatives of LCs and accordingly does not affect our proposed method, as also demonstrated by \citet{Kouzuma2023-ApJ}. 
This arises from the fact that the derivative of a constant term is zero. 
Nevertheless, if the third body is sufficiently bright and exhibits significant variability, it may affect the shape of the derivatives of the LC. 
In such cases, the accuracy of our estimation method could be affected. 

Another possible factor is the presence of starspots. 
However, if we can assume that the brightness of starspots is roughly constant or exhibits little temporal variation, it should not significantly affect our method because of the same reasons mentioned above. 
Indeed, \citet{Poro2024-AJ} investigated the effect of a starspot on mass ratio estimates derived from LC derivatives using the method of \citet{Kouzuma2023-ApJ}. 
They found that the difference in the estimated mass ratios was less than 2\% in each instance. 
Although the estimated parameters differ, both (i.e., mass ratio and inclination) estimation methods rely on the derivative of a LC, so its impact on the inclination estimates is also expected to be similarly very small. 
Nonetheless, if the starspots were to affect the LC enough to change the relative depths of the minima and thus swap the primary and secondary minima, the situation may change. 
For SPp- and SPf-type LCs, it is crucial to obtain the $W_i$ value using the times around the primary eclipse. 
Therefore, if an eclipse that is intrinsically the secondary (shallower) minimum is mistaken for the primary (deeper) minimum, the inclination estimated by our method may differ significantly from the true value.

\section{Summary}\label{sec:summary}
We have proposed a new simple method for estimating the orbital inclinations of overcontact binary systems. 
Essential steps of this method involve classifying the derivatives of the LC following \Kcs (namely, \cite{Kouzuma2025-PASJ}) and extracting key values from these derivatives. 
The key values ($W_i$) are obtained by measuring the time intervals for DP-, SPp-, and SPf-type LCs and the normalized second derivative of the LC at phase 0.5 for SPb- and SPs-type LCs. 
By substituting the obtained key values into the empirical formulae we proposed, their orbital inclinations can be estimated, together with associated uncertainties. 
An application to real binary data demonstrates that the uncertainties estimated by our method are reasonable, as they are comparable to the standard uncertainties. 
The proposed method is expected to be applicable to the LC of any overcontact system. 

A disadvantage of our method is that it requires LCs with sufficiently high photometric precision so that their derivatives do not become excessively noisy. 
One way to achieve this is by improving the photometric precision, or alternatively, by acquiring LCs spanning multiple phases and averaging them to obtain a representative LC. 
However, the photometric accuracy and precision of recent surveys have steadily improved over the years, resulting in an increasing number of highly precise LCs for a large number of eclipsing binaries. 
Accordingly, the above concern is likely to become less significant in the future. 

The computational cost of our method is extremely low, because no iterative process is required. 
In other words, the orbital inclination can be readily estimated once a LC is obtained. 
This would be useful for gaining a preliminary understanding of a system's properties. 
A further advantage is that it facilitates the setting of initial parameters necessary for LC analysis. 
Furthermore, our method allows us to estimate the inclinations of an enormous number of overcontact binaries with trustworthy uncertainties. 
This is also valuable for statistical studies.

\begin{ack}
The author would like to thank the anonymous referee for helpful comments and suggestions that improved the paper. 
This work was supported by JSPS KAKENHI Grant Number 25K07358. 
This paper includes data collected by the TESS mission. Funding for the TESS mission is provided by the NASA’s Science Mission Directorate. 
This paper includes data collected by the Kepler mission and obtained from the MAST data archive at the Space Telescope Science Institute (STScI). 
Funding for the Kepler mission is provided by the NASA Science Mission Directorate. 
STScI is operated by the Association of Universities for Research in Astronomy, Inc., under NASA contract NAS 5–26555. 
This research made use of Lightkurve, a Python package for Kepler and TESS data analysis (Lightkurve Collaboration, 2018). 
\end{ack}

\end{document}